\documentclass[sigplan,screen,dvipsnames]{acmart}
\pdfoutput=1
\usepackage[english]{babel}
\usepackage[utf8]{inputenc}
\usepackage{graphicx,amsthm,url,tikz,float}
\usepackage{amsmath}
\usepackage{caption}
\usepackage{subcaption}
\usepackage[final]{listings}
\usepackage{hyperref}
\usepackage{stmaryrd}
\usepackage{extarrows}
\usepackage{natbib}
\usepackage{booktabs}
\usepackage{dcolumn}
\usepackage{siunitx}
\usepackage{mathpartir}
\usepackage{eso-pic}
\usepackage{ifthen}
\usepackage{tikz}
\usepackage{pgfplots}
\usepackage{xparse}
\usepackage[c1]{optidef}
\usepackage{enumitem}
\usepackage{flushend}
\usepackage{microtype}

\pgfplotsset{compat=1.17}

\usetikzlibrary{arrows.meta}
\usetikzlibrary{svg.path}

\definecolor{dark-green}{rgb}{0,0.5,0}

\newcommand{\IR}[0]{\textbf{IR}}

\title{Fusing Gathers with Integer Linear Programming}

\makeatletter
\DeclareFontFamily{T1}{lmtt}{\hyphenchar \font\m@ne}
\DeclareFontShape{T1}{lmtt}{l}{n}{<->ec-lmtl10}{}
\DeclareFontShape{T1}{lmtt}{l}{it}{<->sub*lmtt/l/sl}{}
\DeclareFontShape{T1}{lmtt}{l}{sl}{<->ec-lmtlo10}{}
\makeatother
\newcommand{\codefamily}{\ttfamily\selectfont}
\lstdefinestyle{haskell}{%
    language=Haskell,
    upquote=true,
    deletekeywords={case,class,data,newtype,default,deriving,do,in,instance,let,of,type,where,if,then,else},
    morekeywords={[2]class,data,newtype,default,deriving,family,instance,type,where,pattern},
    morekeywords={[3]in,out,let,case,of,do,if,then,else},
    literate=
        {\\\\}{{\char`\\\char`\\}}1
        {>->}{>->}3
        {>>=}{>>=}3
        {->}{{$\rightarrow$}}2
        {>=}{{$\geq$}}2
        {<-}{{$\leftarrow$}}2
        {<=}{{$\leq$}}2
        {=>}{{$\Rightarrow$}}2
        {|}{{$\mid$}}1
        {~}{{$\sim$}}1
        {forall}{{$\forall$}}1
        {exists}{{$\exists$}}1
        {...}{{$\cdots$}}3
}
\lstdefinestyle{inline}{
    basicstyle=\small\ttfamily,
    keywordstyle=[1],codefamily
    keywordstyle=[2],
    keywordstyle=[3],
    keywordstyle=[4],
}
\lstdefinestyle{footnote}{
    basicstyle=\footnotesize\ttfamily,
    keywordstyle=[1],
    keywordstyle=[2],
    keywordstyle=[3],
    keywordstyle=[4],
}
\lstdefinestyle{c}{%
    language=C,
    upquote=true,
    keywordstyle=[1]\color{Bittersweet},
    keywordstyle=[2]\color{ForestGreen},
    keywordstyle=[3]\color{Bittersweet},
    keywordstyle=[4]\color{RoyalPurple},
    morekeywords={[3]in}
}
\lstnewenvironment{haskell}
  {\lstset{style=haskell}}
  {}

\lstdefinestyle{python}{%
    language=Python,
    upquote=true,
    deletekeywords={is},
    morekeywords={[2]map,permute,generate,fold,foldSeg,gather,backpermute,scanl,scanr,force,scatter,size,!,zip,zipWith,zipWith3,unzip},
    keywordstyle=[1]\color{Bittersweet},
    keywordstyle=[2]\color{ForestGreen},
    keywordstyle=[3]\color{Bittersweet},
    keywordstyle=[4]\color{RoyalPurple}, 
    literate={lam}{{$\lambda$}}1,
    mathescape
}

\lstnewenvironment{python}
  {\lstset{style=python}}
  {}
\newcommand{\makeatcode}{\lstMakeShortInline[language=python,style=python]@}
\newcommand{\makeatchar}{\lstDeleteShortInline@}

\DeclareMathAlphabet{\mathpzc}{OT1}{pzc}{m}{it}

\definecolor{dblue}{RGB}{100 0 255}
\newcommand{\clus}[2]{[{#1}\mspace{1mu}{|}\mspace{1mu}{#2}]}
\newcommand{\varF}[2]{{\textit{#1}}\mspace{1mu}{:}\mspace{1mu}{#2}}
\newcommand{\varI}[1]{{\textit{#1}}}
\newcommand{\fuse}[3]{{{#1} {\ {\rangle} \ {#2} \ \rangle\!\rangle \ {#3}}}}
\newcommand{\order}[1]{{\mathpzc{#1}}}
\newcommand{\orderL}{\order{l}}
\newcommand{\inv}{\phantom{p{,}}}
\newcommand{\invF}{\phantom{\varF{x}{\order{d}}{,}}}
\newcommand{\sqin}{
  \mathrel{\vphantom{\sqsubset}\text{%
    \mathsurround=0pt
    \ooalign{$\sqsubset$\cr$-$\cr}%
  }}%
}
\newcommand{\defpremise}[1]{x\not\sqin #1 \\\\ } 
\newcommand{\intropremise}[1]{p\not\sqin {#1} \\\\ } 
\newcommand{\glabel}[1]{{\orderL_{#1}}}

\author{David van Balen}
\email{d.p.vanbalen@uu.nl}
\orcid{0000-0002-2807-9860}
\affiliation{%
  \institution{Utrecht University}
  \city{Utrecht}
  \country{Netherlands}
}
\author{Gabriele Keller}
\email{g.k.keller@uu.nl}
\orcid{0000-0003-1442-5387}
\affiliation{%
  \institution{Utrecht University}
  \city{Utrecht}
  \country{Netherlands}
}
\author{Ivo Gabe de Wolff}
\email{i.g.dewolff@uu.nl}
\orcid{0000-0002-4731-2234}
\affiliation{%
\institution{Utrecht University}
\city{Utrecht}
\country{Netherlands}
}
\author{Trevor L. McDonell}
\email{t.l.mcdonell@gmail.com}
\orcid{0000-0001-7806-9751}

\copyrightyear{2024}
\acmYear{2024}
\setcopyright{acmlicensed}
\acmConference[FProPer '24]{Proceedings of the 1st ACM SIGPLAN International Workshop on Functional Programming for Productivity and Performance}{September 6, 2024}{Milan, Italy}
\acmBooktitle{Proceedings of the 1st ACM SIGPLAN International Workshop on Functional Programming for Productivity and Performance (FProPer '24), September 6, 2024, Milan, Italy}
\acmDOI{10.1145/3677997.3678227}
\acmISBN{979-8-4007-1100-8/24/09}
\begin{document}
\begin{abstract}

We present an Integer Linear Programming based approach to finding the optimal fusion strategy for combinator-based
parallel programs. While combinator-based languages or libraries provide a convenient interface for programming parallel
hardware, fusing combinators to more complex operations is essential to achieve the desired performance. Our approach is
not only suitable for languages with the usual map, fold, scan, indexing and scatter operations, but also gather operations,
which access arrays in arbitrary order, and therefore goes beyond the traditional producer-consumer fusion. It can be parametrised
with appropriate cost functions, and is fast enough to be suitable for just-in-time compilation.

\end{abstract}

\maketitle
\makeatcode 
\newtheorem{formulation}{Formulation}

\section{Introduction}

Combinator-based parallel array languages allow programmers to express data parallelism in a high-level way~\cite{dubach,futhark,multicore,Collins2014NOVAAF,gpunesl,copperhead}. They also
expose computation and communication patterns to the compiler, and thereby enable the compiler to exploit these patterns to
generate highly performant code for parallel architectures, such as GPUs and multicore CPUs.  However, to achieve
satisfactory performance for actual programs, it is not sufficient to only provide efficient implementations of all the
built-in combinators such as maps, folds, scans and permutations individually. Instead, it is necessary to combine
and optimise sequences of such combinators. This process is called fusion and in this paper we introduce a novel fusion algorithm, that preserves the parallel structure and finds the best option for fusion, in a larger search space than many existing algorithms.

Already in sequential languages, simply executing the combinators one after the other leads to unnecessary allocation
and traversals of data structures for all but trivial programs. In a parallel setting, it is even more important than in
a sequential setting to minimise data access and allocation by fusing sequences of individual combinators into a few, more
complex operations, since many of these massively data-parallel programs are memory-bound.  This is a well-known issue,
and there is extensive research both in the sequential~\cite{theorypracticeshortcut,stream,unbuildunfoldr,foldrbuild}
and parallel~\cite{delayed,graphred,gpunesl,skelfusion,graphpartit} context on how to fuse individual combinators to
reduce both memory accesses and usage. Unfortunately, most solutions to fusion of sequential programs destroy the
implicit parallelism present in the combinators by transforming them into complex recursive traversals, so we cannot
use these approaches for parallel programs.

To achieve the best possible performance on a parallel architecture, we are essentially interested in partitioning the combinators in an array program into an \emph{optimal} set of fusible clusters. 
Most existing fusion methods are \emph{greedy}, but in the context of data-parallel array combinators, we advocate for an exact solution.
In this paper, we present our approach to finding an optimal loop partitioning that
leverages Integer Linear Programming (ILP) to provide a
\emph{flexible} framework for optimising the program based on a set of
cost metrics, such as minimising the number of manifest arrays or the number of memory
accesses.

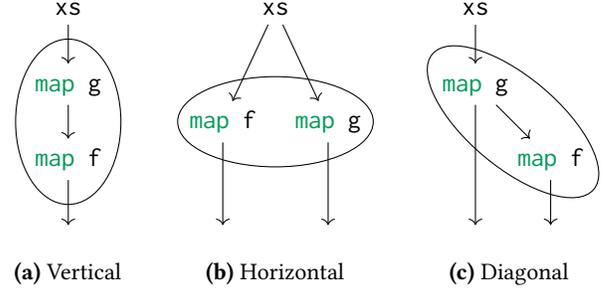
\begin{figure}
    \begin{subfigure}{.1\textwidth}
        \centering
        \begin{tikzpicture}
            \node (xs) at (0,0) {\codefamily xs};
            \node (as) at (0,-1) {\codefamily {\color{ForestGreen}map }g\vphantom{f}};
            \node (bs) at (0, -2) {\codefamily {\color{ForestGreen}map }f\vphantom{g}};
            \node (o1) at (0, -3) {};
            \draw[->]                (xs) -- (as);
            \draw[->]                (as) -- (bs);
            \draw[->]                (bs) -- (o1);
            \draw (0,-1.5) circle [x radius=0.7, y radius=1.1];
        \end{tikzpicture}
        \caption{Vertical}
        \label{fig:vertical}
    \end{subfigure}
    \begin{subfigure}{.2\textwidth}
        \centering
        \begin{tikzpicture}
            \node (xs) at (0,0) {\codefamily xs};
            \node (as) at (-0.7, -1.5) {\codefamily {\color{ForestGreen}map }f\vphantom{g}};
            \node (bs) at (0.7, -1.5) {\codefamily {\color{ForestGreen}map }g\vphantom{f}};
            \node (o1) at (-0.7, -3) {};
            \node (o2) at (0.7, -3) {};
            \draw[->]                (xs) -- (as);
            \draw[->]                (xs) -- (bs);
            \draw[->]                (as) -- (o1);
            \draw[->]                (bs) -- (o2);
            \draw (0,-1.5) circle [x radius=1.3, y radius=0.6];
        \end{tikzpicture}
        \caption{Horizontal}
        \label{fig:horizontal}
    \end{subfigure}
    \begin{subfigure}{.14\textwidth}
        \centering
        \begin{tikzpicture}
            \node (xs) at (0,0) {\codefamily xs};
            \node (as) at (0, -1) {\codefamily {\color{ForestGreen}map }g\vphantom{f}};
            \node (bs) at (1, -2) {\codefamily {\color{ForestGreen}map }f\vphantom{g}};
            \node (o2) at (1, -3) {};
            \node (o1) at (0, -3) {};
            \draw[->]                (xs) -- (as);
            \draw[->]                (as) -- (bs);
            \draw[->]                (as) -- (o1);
            \draw[->]                (bs) -- (o2);
            \begin{scope}[rotate around={140:(0.5,-1.5)}]
                \draw (0.5,-1.5) circle [x radius=1.4, y radius=0.6];
            \end{scope}
        \end{tikzpicture}
        \caption{Diagonal}
        \label{fig:diagonal}
    \end{subfigure}
    \caption{Three types of fusion}
    \label{fig:verhordiag}
\end{figure}

Let us start by looking at different ways to fuse multiple combinators. 
%
The simplest case is a sequence of @map@ operations, which first apply a function @g@ to all elements of an array @xs@,
and then a function @f@ to all elements of the resulting array. 
\begin{python}
map (f, map (g, xs))
\end{python}
Here the inner @map@ produces an array, which is immediately consumed by the subsequent @map@ operation. 
The
well-known map-fusion rule can be used to combine this into a single traversal~@map (lam(x) : f(g(x)), xs)@. This way,
we can eliminate
the intermediate array, as shown in the dependency graph in Figure~\ref{fig:verhordiag}(a), where the arrows represent
fusable edges, and the circles contain the combinators we fuse into one kernel. In this paper, we refer to fusion of a consecutive producer and
consumer as \emph{vertical fusion}~\cite{streamprocessing}.  Now, suppose we need to traverse the same array multiple times, as in the following
example, also visualised in Figure~\ref{fig:verhordiag}(b):
\begin{python}
as = map (f, xs)
bs = map (g, xs)
\end{python}
In contrast to vertical fusion, fusing this program into a single loop will not
reduce the memory usage of the program, but it will reduce the number of memory
accesses that are required.
%
In this paper, we refer to the fusion of two synchronous traversals on the same array
into a single loop as \emph{horizontal fusion}, also known as \emph{tupling}~\cite{Bird80:Tabulation,tuplingChin,streamprocessing}. While
all approaches to fusion can deal with straightforward occurrence of vertical fusion, many local rewrite-based systems~\cite{theorypracticeshortcut,stream,unbuildunfoldr,foldrbuild,delayed} do not handle horizontal fusion.



To illustrate some challenges of fusion, let us consider a more complex program, where it is much less obvious how it can be translated into a single
parallel loop:
\begin{python}
def singleLoop (as):
  bs = reverse (as)
  cs = map (f, as)
  ds = map (g, cs)
  result = zipWith3 (lam(b,c,d) : b+c+d, bs, cs, ds)
  return result
\end{python}

%
As in the first example, the result of @map f@ is consumed by @map g@. 
However, this time, we cannot rewrite this into a single map, since the intermediate array @cs@ is also an input to  @zipWith3@, which sums the corresponding elements of its three input arrays. 
In such a situation, many existing methods take one of two approaches: they either
(1) duplicate work, inlining the computation of @cs@ so it can be fused; or
(2) do not perform fusion.
In many cases, duplicating work is actually better than forgoing fusion, since the
time to access data in main memory is significantly higher than the time to
execute arithmetic operations on values already in registers. However, without
analysing the functions @f@ and @g@ (which may be non-trivial), such work
duplication could result in a slow-down of the program, which is why many fusion
systems conservatively choose to avoid duplicating work.
In this paper we propose a third option:
\emph{diagonal fusion}, shown in Figure~\ref{fig:verhordiag}(c). Diagonal fusion merges the two maps as in vertical fusion, but
stores the result of the intermediate computation as well.
This gives us the best of both worlds: minimising the number of array accesses,
but without introducing work duplication.

Returning to @singleLoop@, there are more facets in this function that complicate fusion: the computation of @bs@ accesses @as@ in a different order than
the computation of @cs@ does, which means that those memory reads cannot be fused horizontally.
We can, however, fuse them into one loop using multiple reads from the same array.

A fusion system that is capable of all these different kinds of fusion is able to fuse @singleLoop@ into a single loop, without any intermediate arrays.
%
There is, however, a challenge associated with supporting all of these fusion techniques at the same time, as it 
may greatly increase the search space. In fact, finding the optimal loop partitioning according to some given
metric is an NP-hard problem~\cite{ILP}. 
In this work, we describe an Integer Linear Programming formulation that models this search space, extending previous work in this domain to add support for order changing operations, like @gather@ and @backpermute@ (backwards permutations).

This paper makes the following contributions:
\begin{itemize}[leftmargin=*]
    \item We present a formalisation of vertical, horizontal, and diagonal fusion,
          and the conditions under which order-changing operations such as @gather@ can fuse
          (\S\ref{sec:fusion});

    \item We present an ILP formulation for the optimal clustering of combinator based 
          parallel array programs, supporting a set of common parallel array combinators
          (\S\ref{sec:ilp});



    \item We present an evaluation of the algorithm, which we have implemented and benchmarked in the Accelerate
      framework~\citep{multicore}, a parallel array language embedded in Haskell (\S\ref{sec:benchmarks}).

\end{itemize}

The rest of this paper is structured as follows:
Section~\ref{sec:language} briefly introduces the language we perform fusion on.
In Section~\ref{sec:fusion}, we discuss fusion semantics and derive rules that describe when certain combinators are fusible,
whereas Section~\ref{sec:ilp} formalises these rules to as Integer Linear Programming problem.
Finally, we evaluate our fusion algorithm in Section~\ref{sec:benchmarks} by looking both at the time required to
find the optimal clustering schedule, as well as the effect of the optimisation on the runtime of the program, for a range of different benchmarks.

\section{Language}\label{sec:language}

We describe clustering as an analysis on a simple array based intermediate language \IR, which is similar to the IR
actually used in Accelerate.
It provides function calls and a small set of basic built-in array combinators, such as @map@, which can be parametrised
with functions on array elements, but not with other array combinators.
We call the top tier \emph{array level} and the expressions passed to combinators \emph{element level}.
\IR\ has one restriction on the order in which combinators are evaluated in (see @scatter@), but is otherwise purely functional.
Arrays and indices in \IR\ are multidimensional, but have a regular hyperrectangular shape and arrays of arrays are not allowed.
\IR\ also
has lambda-abstractions of the form $\lambda$@ x@$_1$@ x@$_2$@ @$\ldots$@ : expr@.

We have the following array combinators and operators in \IR, also listed in Table~\ref{table:singleton-clusters}:

@xs ! n@: The infix operator @!@ is a regular index  operation. It requires its argument array @xs@ to be
fully computed, and therefore prevents operations computing @xs@ to be fused into subsequent operations.
Indexing is the only way to use an array in an element level expression. 
Multidimensional arrays take tuples as their index.

@map (f, xs)@ applies the function @f@ to all elements of @xs@, producing a new array.

@generate (s, f)@ creates a new array of size @s@, initialised with the function @f@ applied to the respective index of the array element.
If @s@ is a tuple, @generate@ creates a multidimensional array.

\lstinline[mathescape]{gather$_\orderL$ (is, as)} takes an array of indices @is@ and a source array @as@. It reads the
elements of @as@ from the source array in the order defined by the index array @is@. Semantically, it is equivalent to the expression
\lstinline[mathescape]{map ($\lambda$i -> as!i, is)}. However, with @gather@, it is explicit that the argument @is@ defines a
traversal order, and @as@ is a parameter, whereas this is not the case for @map@. As we will see later, this extra information allows
us to fuse the computation of the source array @as@ when @gather$_\orderL$@ is used, whereas the @map@ version cannot be
fused.

We can. for example, express the @reverse@ function on arrays in terms of @generate@ and @gather@:
\begin{python}
def reverse (xs):
  n = size (xs)
  inds = generate (n, lami : n-i-1)
  result = gather (inds, xs)
  return result
\end{python}
Every @gather@ in the program is annotated with a unique numerical label $1 < \orderL < g + 2$, where $g$ is the number of gathers in the program. The label is used during the
fusion analysis to refer to the traversal order of the particular @gather@, and the values 0 and 1 are reserved for the normal and reverse order. 
We may omit the label when the label is not important.

@scatter (f, dest, src)@ takes a function @f@, an array @src@ whose elements are pairs @(i, x)@ of a destination index @i@ and a value @x@, and a destination array @dest@. 
For each element @(i, x)@ in @src@, it updates @dest!i@ to @f (src!i, x)@. 
This is a destructive update, which we justify in our intermediate representation by requiring that @scatter@ is the \emph{final} consumer of @dest@: 
After @scatter@, the original destination array is out of scope, and the result of @scatter@ points to the updated version.
In the translation from Accelerate to this intermediate language, the destination array is copied before the @scatter@, but this may not always be required.
The operation is only defined if all @i@ in @src@ are in the range of @size dest@. 
If @op@ is not associative and commutative, and there are multiple elements in @src@ with the same index, the result depends on the order in which the elements of @dest@ are updated, on which no guarantees are given.
@scatter@ is often used as \emph{random writes}, for example in the implementation of @filter@.

@force (xs)@ returns its input array, but prevents fusion of the consumer of @force@ with the producer of
its argument, providing some explicit control over fusion.

@scanl (op, xs)@ and @scanr (op, xs)@ compute the (generalised) prefix sum or cumulative sum of an array with binary operation @op@. For a one-dimensional array, at each index they compute the combined value of all prior (in case of @scanl@) or later (in case of @scanr@) elements.
In a multidimensional array, the scan is performed over each row.
There are multiple ways to implement such scan primitives in data-parallel languages, such as scan-then-propagate~\cite{sengupta2007scan,EfficientParallelScanAlgorithmsForGPUs,vsinkarovs2022parallel}, or a chained scan~\cite{Merrill2016SinglepassPP,wolff2024zero}.
What is important for our purposes is that the @scanl@ operator in \IR~can be computed in a single pass over the data, whether that's a sequential scan per row, the first stage of a three-pass scan, or the chained scan.

We require a three-pass scan to be split before fusion, as in Matsuzaki and Emoto~\cite{parallelskeletons}, because each fused cluster should compile to a single pass over its arrays.

@fold (op, xs)@ performs a reduction. In case of a multi-dimensional input, it performs the reduction over the innermost dimension, and returns an array of one dimension lower. 
Similar to scans, @fold@ should be a single parallel loop, such as a sequential fold per row, or a single stage of a multi-stage data parallel reduction.

@size (xs)@ returns the size of  @xs@.

We do not include a @zip@ or @zipWith@ in \IR, but our surface-level code examples do. 
In our compiler, an Array-of-\\Structures to Structure-of-Arrays transformation happens \emph{before} the fusion analysis.
As a consequence, each array argument to a combinator is actually a set of arrays, 
which means that every @zip@ is a no-op (and every @zipWith@ translates to a @map@) if the two arrays are statically known to have the same size. 
When they are not, in our language, the arrays are first @gather@ed to the appropriate size.
Other languages might assume or enforce that the two arrays have the same size before zipping.
While this is a consequence of our specific internal representation,
our IlP can easily be extended to support an explicit @zip@ combinator.
This would complicate the presentation, because we often refer to arrays by the operation that produces or consumes it.

\subsection{On the Choice of Combinators}\label{ssec:gatherbackpermute_scatterpermute}
We chose a set of combinators for \IR\ which are powerful enough to implement common data parallel patterns in a way that
maximises the opportunities for fusion. 
For example, most data-parallel array combinator languages have a \emph{random reads} combinator and a \emph{random writes} combinator, but there are varying flavours in each of these categories.
We chose @gather (is, xs)@, which takes an array of indices as its argument. 
An alternative version, @backpermute (sz,f,xs)@, takes the size of the new array and an index permutation function as arguments instead.
With our fusion system, we find that @backpermute@ can be implemented in terms of @generate@ and @gather@ without losing any fusibility, but implementing @gather@ in terms of @backpermute@ requires indexing, which (in our system) prevents the @backpermute@ from fusing with the producer of @is@:

\begin{python}
def backpermute (sz,f,xs):
  is = generate (sz,f)
  ys = gather (is,xs)
  return ys
\end{python}
\begin{python}
def gather (is,xs):
  sz = size (is)
  f  = lami : is!i
  ys = backpermute (sz,f,xs)
  return ys
\end{python}
In other words, @backpermute@ is like @gather@, except it forces the user to use the @generate@ interface to create the indices.
The @gather@ interface gives the option to create the indices in a different way, and when they \emph{are} created by a @generate@, it fuses with them.
Whenever we use @backpermute@ in an example in this paper, we refer to the above implementation in terms of @gather@.

Similarly, the version of @scatter@ that we use takes an array with values and target indices as an argument.
We choose to have these values and indices zipped, to ensure that they have the same size.
An alternative version, @permute@, takes an index permutation function as one of its arguments, and again @permute@ can be defined in terms of @scatter@ and @generate@ without losing fusibility or performance.
A common use case for the random write combinator is the definition of @filter@ (or @compact@), and using @scatter@ lets us fuse @filter@ into fewer clusters than @permute@ does.
A similar combinator, where the combination function is always @lamx y. x@, is also often called @scatter@.

The approach in this paper can also be applied to languages wth @backpermute@ and @permute@,
but doesn't fuse across any indexing operations such as the ones they often use.

Finally, we consider expressive power of our combinators.
@map@ and @gather@ can both be defined in terms of @generate@ with explicit indexing, 
and a program containing \emph{only} this function would compile down to the same concrete loop.
However, as we will see in the rest of this paper, the \emph{static constraints} imposed by these more restrictive combinators are exactly what we leverage in order to safely fuse programs.
Of course, @map@ and @gather@ are not the only possible special cases.
It is possible to define many subsets of @gather@, such as a combinator @take (n,xs)@ which returns the first @n@ elements of @xs@.
This combinator would be useful to more accurately describe the behaviour of @zip@, for example, and in turn allow us to perform fusion on clusters that we can't prove to be safe using only the guarantees of @gather@.
We believe that the list of combinators we discuss are reasonable, and the techniques described in this paper can easily be extended to support additional combinators like @take@.

\section{Fusion}\label{sec:fusion}

Our fusion groups those array computations which can be executed together in one single loop into clusters.
In this section, we describe how the different array computations in such a cluster can be fused. 
Each cluster is either a single operation, or a fused combinaton of two clusters. 
The different arrays in a cluster may be fused differently, e.g.\ some may be fused horizontally, and some may not be fused at all.
This is made explicit in the merge operation, which formalizes the ways two clusters can be combined based on the data flow.
In Section~\ref{sec:ilp}, we formulate ILP constraints that exactly match the options described here.
A program transformation to accomplish this fusion analysis, out of scope for this paper, may also be defined in terms of such binary merges.

\subsection{Cluster Notation for Combinators}\label{ssec:clusternotation}
We characterise clusters $\clus{I}{O}$ in this context solely in terms of
the set of their input arrays $I$ and output arrays $O$, and the order in which these are accessed and produced. $I$ and $O$ contain elements $p$ of the form $x$ or $\varF{x}{\order{d}}$.
The annotation $\order{d}$ refers to the order in which the input is consumed or the output produced. The order
$\order{d}$ can either be $0$, for left-to-right traversal, $1$ for right-to-left traversal, or $\order{d} \in \{2..g+1\}$ (with
$g$ the number of @gather@ operations in the program) if the order is determined by a @gather@ operation with
the label $\order{d}$. Inputs or outputs for which the travelsal order is unknown have no annotation.
In \IR, this is the case for array indexing and scatter, which respectively consume and produce an array
in an unpredictable order. The fusion rules, which we introduce later, only fuse arrays with an order annotation.
Some combinators, gather and scans in \IR, put requirements on this order as we will later explain.

For example, @y = map (+1) x@ has $x$ as input and $y$ as output.
It produces the output in the same order as it consumes the input, hence we represent it as
$\clus{\varF{x}{\order{d}}}{\varF{y}{\order{d}}}$. Now consider @y = map (\i -> z ! i) x@, which reads from array $z$ in an unpredictable order.
Hence this is represented as $\clus{\varF{x}{\order{d}},\varI{z}}{\varF{y}{\order{d}}}$.

We give the cluster representation of all combinators in Table~\ref{table:singleton-clusters}. The free array variables
of a term $t$ are denoted by $\textrm{free}\ t$ (e.g., the variable $z$ in the previous @map@ example). Concretely,
these variables are the array variables used as source for array indexing in expressions. As they are accessed in an
arbitrary order, we can not fuse their computations.

\begin{table}
\caption{Input and output of combinators in \IR.}
\centering
{
\setlength{\tabcolsep}{0.1em}
\begin{tabular}{l r c l}
  \hline
  \multicolumn{1}{c}{\textbf{Combinator}} & \multicolumn{2}{r}{\textbf{Cluster}} & \\
  \hline
  \lstinline[mathescape]|ys = generate (s, f)|
    & $[\textrm{free}\ s, \textrm{free}\ f$ & $|$ & $\varF{ys}{\order{d}}]$ \\
  \lstinline[mathescape]|ys = map (f, xs)|
    & $[\textrm{free}\ f, \varF{xs}{\order{d}}$ & $|$ & $\varF{ys}{\order{d}}]$ \\
  \lstinline[mathescape]|ys = gather$_\orderL$ (is, xs)|
    & $[\varF{is}{\order{d}}, \varF{xs}{\orderL}$ & $|$ & $\varF{ys}{\order{d}}]$ \\
    \lstinline[mathescape]|ys = scatter|
      & \ $[\textrm{free}\ op, \varF{src}{\order{d}}, \varI{dest}$ & $|$ & $\varI{ys}]$ \\[-0.2em]
    \lstinline|      (op, dest, src)|& \multicolumn{3}{r}{where $\order{d} \in \{ 0, 1\} $} \\
    \lstinline[mathescape]|ys = fold (op, xs)|
      & $[\textrm{free}\ op, \varF{xs}{\order{d}}$ & $|$ & $\varF{ys}{\order{d}}]$ \\
    \lstinline[mathescape]|ys = scanl (op, xs)|
      & $[\textrm{free}\ op, \varF{xs}{0}$ & $|$ & $\varF{ys}{0}]$ \\
    \lstinline[mathescape]|ys = scanr (op, xs)|
      & $[\textrm{free}\ op, \varF{xs}{1}$ & $|$ & $\varF{ys}{1}]$ \\
    \lstinline[mathescape]|ys = force (xs)|
      & $[\varI{xs}$ & $|$ & $\varI{ys}]$ \\
  \hline
\end{tabular}
}
\label{table:singleton-clusters}
\end{table}

\subsection{Fusion Inference Rules}\label{sec:inference}
\begin{figure}
\small
{
\makeatletter
\setlength{\mpr@andskip}{1em}
\begin{mathpar}
\inferrule[Vertical]      {\defpremise{I_2}\fuse{\clus{I_1}{\invF{}O_1}}{\clus{\invF{}I_2}{O_2}}{\clus{I}{O}}}
                          {\fuse{\clus{I_1}{  \varF{x}{\order{d}}{,}O_1}}{\clus{  \varF{x}{\order{d}}{,}I_2}{O_2}}{\clus{I}{O}}} \and
\inferrule[Horizontal]    {\fuse{\clus{\invF{}I_1}{O_1}}{\clus{\invF{}I_2}{O_2}}{\clus{\invF{}I}{O}}}
                          {\fuse{\clus{ \varF{x}{\order{d}}{,}I_1}{O_1}}{\clus{ \varF{x}{\order{d}}{,}I_2}{O_2}}{\clus{ \varF{x}{\order{d}}{,}I}{O}}} \\
\inferrule[Diagonal]      {\defpremise{I_2}\fuse{\clus{I_1}{\invF{}O_1}}{\clus{\invF{}I_2}{O_2}}{\clus{I}{\invF{}O}}}
                          {\fuse{\clus{I_1}{ \varF{x}{\order{d}}{,}O_1}}{\clus{ \varF{x}{\order{d}}{,}I_2}{O_2}}{\clus{I}{ \varF{x}{\order{d}}{,}O}}}   \and
\inferrule[Empty]{\phantom{a}}
                          {\fuse{\clus{\varnothing}{\varnothing}}{\clus{\varnothing}{\varnothing}}{\clus{\varnothing}{\varnothing}}} \hspace{20pt} \\ 
\inferrule[IntroI1]       {\intropremise{O_2}\fuse{\clus{\inv{}I_1}{O_1}}{\clus{I_2}{O_2}}{\clus{\inv{}I}{O}}}
                          {\fuse{\clus{  p{,}I_1}{O_1}}{\clus{I_2}{O_2}}{\clus{  p{,}I}{O}}} \and
\inferrule[IntroI2]       {\intropremise{O_1}\fuse{\clus{I_1}{O_1}}{\clus{\inv{}I_2}{O_2}}{\clus{\inv{}I}{O}}}
                          {\fuse{\clus{I_1}{O_1}}{\clus{  p{,}I_2}{O_2}}{\clus{  p{,}I}{O}}} \\
\inferrule[IntroO1]       {\intropremise{I_2}\fuse{\clus{I_1}{\inv{}O_1}}{\clus{I_2}{O_2}}{\clus{I}{\inv{}O}}}
                          {\fuse{\clus{I_1}{  p{,}O_1}}{\clus{I_2}{O_2}}{\clus{I}{  p{,}O}}} \and
\inferrule[IntroO2]       {\intropremise{I_1}\fuse{\clus{I_1}{O_1}}{\clus{I_2}{\inv{}O_2}}{\clus{I}{\inv{}O}}}
                          {\fuse{\clus{I_1}{O_1}}{\clus{I_2}{  p{,}O_2}}{\clus{I}{  p{,}O}}} \\
\end{mathpar}
\makeatother
}
\caption{Inference rules for fusing clusters}
\label{fig:inferencerules}
\end{figure}
Based on the cluster representation we can now formalise how two clusters can be merged into a larger cluster.
The judgement:
\[{\fuse{\clus{I_1}{O_1}}{\clus{I_2}{O_2}}{\clus{I}{O}}}\]
states that fusing cluster $\clus{I_1}{O_1}$ with $\clus{I_2}{O_2}$ results in the new cluster $\clus{I}{O}$.

We illustrate the rules listed in Figure~\ref{fig:inferencerules} with an example. Consider the fusion of clusters
$\clus{\varF{x}{\order{d}}}{\varF{y}{\order{d}}}$ and $\clus{
  \varF{x}{\order{d}}{,}\varF{y}{\order{d}}}{\varF{z}{\order{d}}}$, corresponding to
@y = map f x@ and @z = zipWith g x y@.
They can be fused in a single cluster represented as $\clus{\varF{x}{\order{d}}}{\varF{z}{\order{d}}}$, where
$x$ is shared horizontally, $y$ is fused vertically, and $z$ via \textsc{IntroO2}.
All three arrays are consumed and produced in the same order $\order{d}$.



Our fusion rules are not symmetric: vertical and diagonal fusion can only occur
between an output of the first cluster and an input of the second cluster. 
This asymmetry reflects the behaviour of fusing imperative loops by concatenating their bodies: 
the top half of the resulting loop cannot use results produced by the bottom half.


The inductive rules describe different fusion classes: 
\begin{itemize}[leftmargin=*]
    \item \textsc{Vertical} describes one output $x$ from $O_1$ being streamed into an input $x$ from $I_2$.
    Note that the argument set of the resulting cluster has no mention of $x$: it is completely fused away.
    \item \textsc{Horizontal} describes the input $x$ being shared between the two clusters. 
    The resulting cluster only reads $x$ once. 
    \item \textsc{Diagonal} mirrors \textsc{Vertical}, just like its informal description did. 
    The only difference is the resulting argument list: $x$ is preserved, not fused away.
\item \textsc{Empty} covers the base case, the fusion of two empty clusters.
    \item The remaining \textsc{Intro} cases are required for the trivial independent fusion where one cluster has an input or output argument that is completely independent of the other cluster. 
    This argument is preserved by fusion.
\end{itemize}
In some of these rules we have extra constraints on the absence of the array we add in the sets in the premise.
For \textsc{Vertical} and \textsc{Diagonal}, these constraints prevent us from fusing over an array when it is also read from in a different order.
For the \textsc{Intro} rules, we are preventing one array from being both the input and the output of a single cluster.
We use the notations $x \not \sqin A$ and $\varF{x}{\order{d}} \not \sqin A$ to mean $\forall{\order{d}_1}.\ x,\varF{x}{\order{d}_1} \not \in A$.
The given constraints are sufficient to show that $x \not \sqin$ any of the sets in the premises, assuming that no array can ever be both the input and output of the same cluster.


\subsection{Traversal-Order Sensitive Fusion}\label{sec:gatherrules}
As we can see in Table~\ref{table:singleton-clusters}, some combinators in \IR, like @map@, can take arrays of any traversal
order $\order{d}$ as input, and produce the output in the same order. Others impose a particular order, such as @scanl@, @scanr@, and @gather@
on its second argument.
Horizontal, vertical and diagonal fusion rules are only applicable if the orders match. 



For instance, @y = scanl (+) x@ cannot be fused horizontally with @z = scanr (+) x@: these operations are represented as
singleton clusters $\clus{\varF{x}{0}}{\varF{y}{0}}$
and $\clus{\varF{x}{1}}{\varF{z}{1}}$, and the rule for horizontal fusion is not applicable here.
They can, however, be fused independently (as two separate traversals over $x$ in a single loop).

Let us now look at how @gather@ affects fusion, in which cases it can be fused, and what happens to the iteration
structure. Vertically or diagonally fusing @gather@ \emph{before} other combinators is similar to any other combinator,
but fusing vertically or diagonally \emph{after} another combinator, or fusing horizontally, is more complex, as we will show
in the following examples.



Consider the function @simple1@ below. We can generate a vertically fused loop computing @bs@, by pushing the index permutation through the @map@.
On the left we show the program in our combinator-based intermediate language,
and on the right the corresponding imperative loop (nest).

\vspace{-10pt}\noindent\begin{minipage}[t][][t]{.35\linewidth}%
\begin{python}
def simple1 (is,xs):
  as = map (f, xs)
  bs = gather$_\orderL$ (is, as)
  return bs
\end{python}
\end{minipage}%
\hfill%
\noindent\begin{minipage}[t][][t]{.5\linewidth}%
\begin{lstlisting}[style=c]
for j in 0..is.size {
  i = is[j];
  x = xs[i];
  a = f(x);
  b = a;
  bs[j] = b; }
\end{lstlisting}
\end{minipage}\\
In our notation, this cluster is represented as $\clus{\varF{\textit{is}}{0},
\varF{\textit{xs}}{\orderL}}{\varF{\textit{bs}}{0}}$, choosing the order of @is@ to be $0$.
By fusing the two computations into one loop, work might be duplicated, if  @is@ is larger than
\textit{xs}. However, it may also save work if it is the other way around.

This is an inherent effect of fusing @gather@-like combinators, where the elements of an input array may be used
multiple times, or not at all. 
In this paper, we ignore this difference, as not all array sizes are known at compile time.

Fusion also works on two subsequent @gather@s:

\vspace{-10pt}\noindent\begin{minipage}[t][][t]{.45\linewidth}%
\begin{python}
def simple2 (is1, is2, xs):
  as = gather$_\glabel{1}$ (is1, xs)
  bs = gather$_\glabel{2}$ (is2, as)
  return bs
\end{python}
\end{minipage}%
\hfill%
\noindent\begin{minipage}[t][][t]{.4\linewidth}%
\begin{lstlisting}[style=c]
for i in
    0..is2.size {
  i2 = is2[i];
  i1 = is1[i2];
  x  = xs[i1];
  a  = x;
  b  = a;
  bs[i] = b; }
\end{lstlisting}
\end{minipage}\\
This is represented as $\clus{\varF{\textit{is2}}{0}, \varF{\textit{is1}}{\orderL_2}, \varF{\textit{xs}}{\orderL_1}}{\varF{\textit{bs}}{0}}$.
Note that in @simple2@, we compute the indices in reverse order, before computing the combinators in their normal order.

Now consider the function @simple3@, which only differs from @simple1@ in that the intermediate array @as@ is now part
of the result, and cannot be fused away.  This looks like a classical example of \emph{diagonal fusion}. Looking at it
more carefully, we can see that we cannot fuse it into a single loop:
There is no guarantee that @is@ contains all indices of @as@, which means that the naive attempt below may not compute all elements of @as@.

\vspace{-10pt}\noindent\begin{minipage}[t][][t]{.3\linewidth}%
\begin{python}
def simple3 (is, xs):
  as = map (f, xs)
  bs = gather$_\orderL$ (is, as)
  return (as, bs)
\end{python}
\end{minipage}%
\hfill%
\noindent\begin{minipage}[t][][t]{.5\linewidth}%
\begin{lstlisting}[style=c]
for j in 0..is.size {
  i = is[j];
  x = xs[i];
  a = f(x);
  b = a;
  as[i] = a;
  bs[j] = b; }
\end{lstlisting}
\end{minipage}\\
The corresponding cluster is $\clus{\varF{\textit{is}}{0}, \varF{\textit{xs}}{\orderL}}{\varF{\textit{as}}{\orderL},
  \varF{\textit{bs}}{0}}$.

We prevent this problem by putting one additional restriction on the top-level clusters: 
we require all their output arrays to have order zero or one, guaranteeing that they are fully evaluated.
This excludes the clustering above, as @as@ has the order $\orderL$.

A @gather@'s index permutation can be pushed into @generate@ or through another @gather@, just as done with @map@. 
The next example demonstrates that we can even do it
with a multidimensional @fold@ (which folds over one dimension of the array):

\vspace{-10pt}\noindent\begin{minipage}[t][][t]{.4\linewidth}%
\begin{python}
def simple4 (is, xs):
  m,n = size xs
  as = fold (f, xs)
  bs = gather$_\orderL$ (is, as)
  return bs
\end{python}
\end{minipage}%
\hfill%
\noindent\begin{minipage}[t][][t]{.5\linewidth}%
\begin{lstlisting}[style=c]
for i in 0..m {
  i2 = is[i];
  i3 = i2 * n;
  a  = xs[i3];
  for j in 1..n {
    x = xs[i3+j];
    a = f(a,x); }
  b     = a;
  bs[i] = b; }
\end{lstlisting}
\end{minipage}\\
Here, @gather@ computes the output index for the @fold@, and
then @fold@ loops over the corresponding input range.
The fold cluster is represented as $\clus{\varF{\textit{xs}}{\order{d}}}{\varF{\textit{as}}{\order{d}}}$ and the gather as $\clus{\varF{\textit{is}}{\order{e}}, \varF{\textit{as}}{\orderL}}{\varF{\textit{bs}}{\order{e}}}$. By choosing $\order{d}$ as $\orderL$ and $\order{e}$ as $0$, this can be fused as $\clus{\varF{\textit{xs}}{\orderL}}{\varF{\textit{bs}}{0}}$.

A @scan@, on the other hand, cannot be fused vertically before a @gather@, because the inherent data dependencies are present in the output too.

Finally, we have \emph{horizontal fusion}, as in @simple5@, where @as@ and @bs@ are accessing @xs@ in a different \emph{order}, which means that they cannot share the array access.
This does not mean that the resulting loops cannot be fused, because we can still compile it into a single loop by reading from @xs@ twice.
A cost function should not treat this as horizontal fusion, because the array accesses are not shared.
To avoid work duplication, they cannot both be fused with the producer of @xs@.

\vspace{-10pt}\noindent\begin{minipage}[t][][t]{.44\linewidth}%
\begin{python}[mathescape]
def simple5 (is, xs):
  as = map (g, xs)
  bs = gather$_\orderL$ (is, xs)
  cs = zipWith (h, as, bs)
  in cs
\end{python}
\end{minipage}%
\hfill%
\noindent\begin{minipage}[t][][t]{.45\linewidth}%
\begin{lstlisting}[style=c]
for i in 0..min(
    is.size, xs.size) {
  x  = xs[i];
  a  = g(x);
  i1 = is[i];
  b  = xs[i1];
  c  = h(a,b);
  cs[i] = c; }
\end{lstlisting}
\end{minipage}\\
This cluster traverses $xs$ in order zero, left-to-right, and $\orderL$, the order of to the @gather@ operation, and is thus represented as $\clus{\varF{xs}{0},\varF{xs}{\orderL},\varF{is}{0}}{\varF{cs}{0}}$.

In summary, we've seen that vertically fusing combinators \emph{before} a
@gather@ involves pushing the index permutation function through those
combinators, which is not always possible (c.f.\ @scan@).
In order to model the legal clusterings, and correctly evaluate their cost, we need to consider the \emph{order} in which an array is accessed.
This order variable is either zero (left-to-right), one (right-to-left, to accommodate @scanr@), or the unique label of a @gather@ in the program.
Note that each @gather@ may not use the same array of indices: thus, we can not
just add one case for the `gathered order', but rather we need a unique label
for each @gather@.
Even when it is possible to fuse a @gather@, doing so naively sometimes causes us to only partially evaluate other arrays, which is something we need to avoid when these other arrays have other outgoing edges in the data dependence graph. This is prevented by requiring that all output variables of a cluster have order zero or one: their order may not depend on a @gather@.

Section~\ref{sec:gatherilp} describes the formalisation of these rules in our ILP formulation.

\subsection{Optimal Clustering}
In the examples we looked at so far, the optimal fusion outcome was clear, as it was always possible to reduce the program to one
single traversal. 
This is different in the program below:
\begin{python}
def scatterExample (xs):
  as = map (lamx : (x*2, x), xs)
  bs = map (+1, xs)
  result = scatter (+, bs, as)
  return result
\end{python}
Here, the array @bs@ has to be fully computed in a loop before @scatter@ can start. 
This is reflected by the representation of @scatter@, $\clus{\varF{as}{0},\varI{bs}}{\varI{result}}$: 
variable $bs$ is not annotated with an order, hence it cannot be fused.
It is possible to fuse the computation of @as@ with @bs@, as they both
iterate over the same array, and fusing these loops would eliminate one read for each element of @xs@.
Alternatively, we can fuse the computation of @as@ into the @scatter@ computation, and thereby save a read
and a write for each element of @as@ while eliminating the intermediate array entirely. In this example, we only have two
options, and the second is clearly preferable, but in general finding the optimal partitioning is NP-hard~\cite{ILP}.


\begin{figure}[t]
    \begin{tikzpicture}
        \node (xs1)                               {xs\vphantom{abxresult}};
        \node (as1)  [below of=xs1, left  of=xs1] {as\vphantom{abxresult}};
        \node (bs1)  [below of=xs1, right of=xs1] {bs\vphantom{abxresult}};
        \node (res1) [below of=as1, right of=as1] {result\vphantom{abxresult}};
        \draw[-{Implies},double]                (xs1) -- (as1);
        \draw[-{Implies},double]                (xs1) -- (bs1);
        \draw[->]                (as1) -- (res1);
        \draw[-{Implies},double] (bs1) -- (res1);
        
        \draw (0, -1.02) circle [x radius=1.3, y radius=0.3];
        \draw (0, -2.02) circle [x radius=0.6, y radius=0.25];
        
        \node (spacing1) [right of=xs1] {};
        \node (spacing2) [right of=spacing1] {};
        \node (spacing3) [right of=spacing2] {};
        
        \node (xs2)  [right of=spacing3]          {xs\vphantom{abxresult}};
        \node (as2)  [below of=xs2, left  of=xs2] {as\vphantom{abxresult}};
        \node (bs2)  [below of=xs2, right of=xs2] {bs\vphantom{abxresult}};
        \node (res2) [below of=as2, right of=as2] {result\vphantom{abxresult}};
        \draw[-{Implies},double]                (xs2) -- (as2);
        \draw[-{Implies},double]                (xs2) -- (bs2);
        \draw[->]                (as2) -- (res2);
        \draw[-{Implies},double] (bs2) -- (res2);
        
        \begin{scope}[rotate around={135:(3.6,-1.5)}]
            \draw (3.6,-1.5) circle [x radius=1.1, y radius=0.7];
        \end{scope}
        \draw (5, -1) circle [x radius=0.3, y radius=0.3];
    \end{tikzpicture}
    \caption{Possible clusterings of the function \lstinline{scatterExample}}\label{fig:clusterings}
\end{figure}
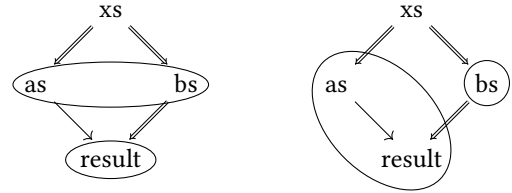

Fusion is the process of placing the array operations in an ordered sequence of clusters. A cluster is defined as
previously discussed, and the sequence of clusters has to adhere to the following rules:
\begin{itemize}
  \item For each cluster, all orders on output arrays have to be either zero or one. This ensures that every element is computed.
  \item For each merge of two clusters, at least one array has to be fused via \textsc{Horizontal}, \textsc{Vertical} or \textsc{Diagonal}. This ensures that we don't fuse loops with different iteration sizes.
  \item The relative order among operations needs to be preserved: if operation $a$ has $xs$ as output and $b$ has $xs$
    as input, then the cluster of $a$ has to come before, or be the same as, the cluster of $b$.
  \item The \emph{temporal uniqueness} of @scatter@ needs to be preserved: If a cluster $a$ contains 
    @scatter(op, src, dest)@, then all other uses of @dest@ have to come before cluster $a$.
\end{itemize}

If the array program is a simple sequence of array combinators, without conditionals and loops, then ``comes before'' means that one operation occurs before the other operation in that sequence.
If the program does have control flow, then the ``comes before'' relation is given by the control flow graph.

Figure~\ref{fig:clusterings} shows the data dependence graph of \\@scatterExample@, and two possible clusterings. Every node
represents either the result of the application of an array combinator (denoted by the variable name its result is bound
to), or a separate graph (see the paragraph on control flow in Section~\ref{ssec:graphandconstraints}). Every edge represents a data
dependency between two nodes, imposed by an array value. Each value may result in multiple edges, all originating from
the node which creates it, pointing at each node that consumes it.  Double arrows indicate an \textit{infusible edge},
that is, a dependecy on a computation which we cannot fuse into the subsequent computation, as the array has to be fully
present before it starts. Aside from @scatter@, the edges from the computation of the input array are also infusible,
since we do not have access to it here. To maximise fusion, the definition of @scatterExample@ can be inlined at its
caller.  Infusible edges can also be caused by indexing into an array, the @force@ combinator, foreign functions or
array-level control flow, such as conditionals and while-loops. Single arrows indicate \emph{fusible} edges: here, @as@
can be computed either in an earlier cluster than @cs@, or in the same cluster as @cs@ (in which case they are fused).

Fusion should find the best clustering, according to some cost metric. In this work, we base the cost on the number of clusters, the number of fused inputs and outputs, the number of intermediate arrays and/or the number of memory reads and writes.
We ignore the sizes and rank of these arrays, for simplicity and because Accelerate does not have static array sizes, but note that these could only improve our cost metrics.

\section{Integer Linear Programming formulation}\label{sec:ilp}
Since our fusion approach offers significantly more options compared to most others, it also means that the
solution space is much larger, and simple heuristics are not sufficient. Therefore, we turn to Integer Linear Programming (ILP) to find the best clustering.
Our Integer Linear Programming formulation is based on existing work by Megiddo~\cite{ILP} and Darte~\cite{darte2002new}. 
We use the original formulation with minor adaptations to support a large part of \IR\ in Section~\ref{ssec:graphandconstraints},
and present our main contribution, the support for @gather@ and similar operations, in Section~\ref{sec:gatherilp}.

\subsection{Data Dependence Graph and Fusion Constraints}\label{ssec:graphandconstraints}

Building the data dependence graph for a program is pretty straightforward: we keep track of the source of each array
in the environment, and whenever it is used by a combinator, we add a fusible or infusible edge between the source label and the consumer
label.
A @force@ does not produce a new node, but adds an infusible edge between its argument and all its consumers.

\paragraph{Graph representation.} A graph $(n, N, F, I)$ is represented by the number of nodes $n$, the set of its node names $N$, fusible
edges $F \subset N \times N$ and infusible edges $I \subset N \times N$. The graphs in Figure~\ref{fig:clusterings} are visual representations of the following graph:
\begin{align*}
  n&=4,\\
  N&=\{\textit{xs}, \textit{as},\textit{bs},\textit{result}\},\\
  F&=\{(\textit{as},\textit{result})\},\\
  I&=\{(\textit{xs},\textit{as}),(\textit{xs},\textit{bs}),(\textit{bs},\textit{result})\}.
\end{align*}

The clustering, that is, assignment of combinators to clusters, of a given graph is visualised using ellipses, and characterised by the mapping $\pi$ of
nodes to cluster number, where $\pi_a \in \mathbb{N}$ for $a \in N$ denotes the number of the cluster that combinator $a$ is assigned to. The clusters are executed in the ascending order of their cluster numbers. 

Additionally, nodes can only be in the same cluster if they are not connected via an infusible edge. 
To express these constraints, we need a second variable type in our cluster characterisation: 
$x_{ab}$ which is $0$ if combinators $a$ and $b$ are fused, i.e. if they are in the same cluster, and $1$ otherwise.
\begin{enumerate}
\item[(3)] $x_{ab}$ is a natural number between $0$ and $1$.
\item[(4)] $x_{ab}$ is $1$ if $(a, b) \in I$.
\item[(5)] $x_{ab}$ is $0$ if $a$ and $b$ are in the same cluster. 
\end{enumerate}
The requirements correspond to the following contraints:
\begin{align}
   {0 \leq \pi_a \leq n}{\quad}{}\\
   {\pi_a \leq \pi_b}{\quad}{(a,b)\in F \cup I}\label{superfluous} \\
   {0 \leq x_{ab}\leq 1}{\quad}{(a,b)\in F}\\
   {x_{ab}=1}{\quad}{(a,b)\in I}\\
   {x_{ab} \leq \pi_b - \pi_a \leq n \cdot x_{ab}}{\quad}{(a,b)\in F \cup I}\label{implies}
\end{align}
Note that Constraint~\ref{superfluous} is implied by Constraint~\ref{implies}, so we do not need to include it in our ILP.\@
These constraints over two categories of variables are enough to translate our data flow graphs into an ILP solution space,
but in order to support our array combinators and intended cost functions, we need more variables and constraints:
The variable $m_a$ denotes whether the array $a$ is completely fused away, or \emph{manifest}.
We choose 0 to mean \emph{manifest}, and 1 to mean \emph{fused away}.
An array can only be fused away if all its outgoing edges are fused over.
\begin{align}
  {x_{ab} \leq 1 - m_a}\\
  {0 \leq m_a \leq 1}
\end{align}
For each @scatter@, we add a constraint that the incoming edge from the \emph{destination} array is infusible (note: not for the \emph{source} array), and we make all outgoing edges (accesses to the destination array after the @scatter@) infusible too.
We also add a constraint that all other uses of the destination array need to be in earlier clusters than the @scatter@,
to accomodate for the destructive update.
\begin{align}
  {x_{ij}=1}{\quad}{j \in \textit{scatter}}\\
  {\pi_i < \pi_j}{\quad}{(d,i)\in F \cup I, j = \textit{scatter(op,src,d)}}
\end{align}

\paragraph{Cost function} 
Next, we need to determine a cost function and express it in terms of constraints, such that the ILP solver can
find the optimal partitioning with respect to this metric. Finding the most suitable cost function for a specific
architecture and purpose is out of scope for this paper. 
Our focus is instead on modelling the set of valid partitionings as an ILP.\@
To demonstrate the versatility of our approach, we have implemented five
different cost functions that
(a) minimise the number of clusters;
(b) maximise the number of fused edges;
(c) maximise the number of intermediate arrays which can be fused; and
(d) minimise the amount of memory read and written (assuming, for simplicity, that each array has the same size). Other
factors, such as the number of registers required or how well the cache is utilised, could also be considered.
For the cost functions we have implemented, this corresponds to:

\begin{enumerate}[leftmargin=*,label=(\alph*)]
 \item Minimize the number of clusters: we add a variable $\pi_{\max}$ and minimize it, adding the constraint:
      \begin{align}
      \forall a. \pi_a \leq \pi_{\max}
      \end{align}

\item Maximising the number of fused edges: Minimise the sum of all $x$ variables.

\item Maximising the number of intermediate arrays removed by fusion, which minimises the number of writes. This means maximising the sum of all $m_a$ variables.

\item Minimise the number of memory reads: This is tricky to measure due to horizontal fusion.
To appropriately measure the number of array accesses, we need to add a number of constraints that is quadratic in the \emph{fusible out-degree} of our data dependence graph.
Note that being quadratic in the \emph{out-degree} is much better than being quadratic in the total size, because in practice programs will rarely have arrays with a large number of outgoing fusible edges. 
These variables group the consumers of each array on the cluster in which the array is consumed, and a variable similar to $\pi_{\max}$ counts the number of times the array is read from. Finally, we minimize sum of these values over all arrays.

\item Minimizing the number of memory reads and writes: Here, we simply minimize the sum of the previous cost function minus the one before it. Any linear combination of valid cost functions is allowed.
\end{enumerate}


\paragraph{Control flow}
The formulation by Megiddo and Sarkar~\cite{ILP} assumes a program without any conditional control flow.
For our language we treat constructs such as array-level conditionals and loops
as a single node that cannot fuse with any other nodes, and construct a separate
ILP for each contained code block, since the fusion choices we make within a branch of a conditional are completely independent of the fusion choices outside of it.
Some extensions to this system, such as keeping track of the number of uses of arrays (compile-time reference counting) to be able to perform certain operations in-place, would require these ILPs to be combined.
This is not a problem, we just need to rename the variables, append the constraints, and sum the cost functions.

\paragraph{Gather}
In the coming subsections, we will see that the current formulation is not yet enough to model the search space for programs with @gather@, and we will add new variables and constraints to address the problems that arise.

\paragraph{Performance and correctness}
Megiddo and Sarkar described two equivalent ILP formulations, a naive version and a simplified one which is linear in the size of the data dependence graph~\cite{ILP}. 
The naive version has a quadratic numer of variables and a cubic number of constraints (in the size of the data dependence graph).

This linear formulation, which we have described above, is much more suitable for a compiler, but it also comes with a
downside, as it provides no way to force all combinators in a cluster to be related: it may result in multiple
completely independent loops placed in one cluster, which might not even have the same iteration size.  This is not
sound, but it can easily be fixed by splitting these clusters after the ILP.
This is done by considering the set of all nodes in the cluster as well as their \emph{parents} in the data flow graph, partitioning that set into \emph{connected components}, and then discarding the added parents again. These parents are required to retain \emph{horizontal fusion}.
Unfortunately, this makes any cost function which uses the \emph{number of clusters} as a measure worse, as it optimises for an imaginary target.
In practice, we do not expect this to be a big problem, as other factors (such as memory) should be much more relevant for predicting the true cost of a partitioning.

We've shown some example linear-sized cost functions above, 
but in situations where an ILP that is quadratic in the size of the data dependence graph is feasible, 
it is possible to take even more factors into consideration.
For example, we could constrain the number of registers that a cluster is allowed to use, or correctly count the number of clusters.
Currently, our ILP solve times are very reasonable (see Section~\ref{sec:benchmarks}), so this bump in complexity might be a reasonable trade-off, especially for applications where programs are small, their runtime is large, or they are run much more often than they are compiled.

To ensure that each cluster has a single iteration size, we use the data dependence graph again, considering only the fusible edges.
We add all the parent nodes of each node in the cluster, separate the resulting graph into connected components, and then remove these parent nodes again.
The combinators in each resulting cluster are all connected by a path of fusible edges, either within the cluster or through a common input, which is enough to ensure that the iteration sizes match up.
The presence of @gather@ complicates this step to some extent: rather than adding each parent node directly, they have
to be be added once for every \emph{order} they are read in.

\paragraph{Possible extensions}
We can easily adapt the cost function to consider other factors which influence the choice of fusion clusters.
For example, duplicating work (instead of storing the result), performing destructive or in-place updates, and re-using memory, are potent optimisation techniques which can be enabled or prevented depending on the partitioning.
This means that ideally, we should not have separate passes for these optimisations before or after our fusion pass, but instead incorporate them into our ILP formulation.

\subsection{Variables and Constraints for Traversal Order}\label{sec:gatherilp}
We can formalise the rules from Section~\ref{sec:gatherrules} using numeric variables $\order{d}^i_a$ and $\order{d}^o_a$ representing the order in which the input and output arrays of combinator $a$ are produced and consumed.
These correspond with the annotations of the form $\varF{x}{\order{d}}$ from the cluster representation
and are either zero, for a left-to-right traversal, one, for right-to-left, or the unique label of a @gather@.

If an array produced by a combinator $a$ is not fused away (that is, $m_a = 0$), then $\order{d}^o_a$ has to be $0$ or $1$, to ensure
that the entire array is filled:
\begin{align}
  {\order{d}^o_a \leq 1 + g \cdot m_a}{\quad}{}
\end{align}
Variable $g$ denotes the number of @gather@s in the program. This constraint allows $\order{d}^o_a$ to be the label of any @gather@ in the program if $m_a$ is one, since each label $\orderL$ is $1 < \orderL < g + 2$.
We also require a constraint between the fusion variable $x_{ab}$, $\order{d}^o_a$, and $\order{d}^i_b$:
If combinators $a$ and $b$ are fused (i.e., $x_{ab} = 0)$, then the order of the output array of $a$ should match the order of the input array of $b$:
\begin{align}
    {-(g+2)\cdot x_{ab}\leq \order{d}^o_a - \order{d}^a_b \leq (g+2) \cdot x_{ab}}{\quad}{(a,b) \in F}
\end{align}
If $a$ and $b$ are not fused, this constraint allows $\order{d}^o_a$ and $\order{d}^a_b$ to be any valid order,
since orders are between zero and $g + 2$ and this constraint allows them to differ by $g+2$.

All that remains to do is relating these variables to the other combinators.
@generate@ can produce an array in any order.
The combinators @scanl@ and @scanr@ can only produce and consume arrays in order
$0$ or $1$, respectively.\@
@map@, @gather@ and @fold@ can produce an array in any order, but we need to compute the order of their input depending on the output.
For @map@, the order of the input is the same as the order of the output.
The second input of a \lstinline[mathescape]{gather$_\orderL$}, the array of indices, is consumed in the same order as its output,
while the first input is consumed in the order $\orderL$, the unique label of the gather.
Finally, for @fold@s, we say that that input order has the same value as the output order.
This stretches the intuition for our order variable, but is sound, and any other choice would prohibit either vertical fusion of a fold into a generate, or diagonal fusion of anything into a fold.

In total, we get the ILP formulation in Figure~\ref{form:final}.
\begin{figure}
\makeatchar
$\begin{array}{@{}l@{}}
\displaystyle \operatorname*{minimize}_{x,\pi,m,\order{d}^o,\order{d}^i}\ \text{\emph{cost function}}\\
\text{subject to} \\[2pt]
\begin{array}{@{}ll@{}}
\pi_a \leq \pi_b                                        & (a,b)\in F \cup I, \\

x_{ab} \leq \pi_b - \pi_a \leq n \cdot x_{ab}           & (a,b)\in F \cup I, \\
x_{ab} \leq 1 - m_a                                     & (a,b)\in F \cup I, \\
0 \leq x_{ab}\leq 1                                     & (a,b)\in F, \\
x_{ab}=1                                                & (a,b)\in I, \\
-(g + 2)\cdot x_{ab}\leq \order{d}^o_a - \order{d}^i_b  & (a,b) \in F, \\
\order{d}^o_a - \order{d}^i_b\leq (g + 2) \cdot x_{ab}  & (a,b) \in F, \\
\order{d}^o_a \leq 1 + g \cdot m_a                      &  \\
0 \leq m_a \leq 1                                       &  \\
0 \leq \pi_a \leq n                                     &  \\
\order{d}^i_a = \order{d}^o_a = 0                       & a \in \textit{scanl}, \\
\order{d}^i_a = \order{d}^o_a = 1                       & a \in \textit{scanr}, \\
\order{d}^i_a = \order{d}^o_a                           & a \in \textit{map} \cup \textit{fold}, \\
\order{d}^i_a = \orderL                                 & (a, \orderL) \in \textit{gather}, \\
x_{ij}=1                                                & j \in \textit{scatter}, \\
\pi_i < \pi_j                                           & j = \textit{scatter}(\hspace{-1pt}\textit{op},\textit{src},d),\\
                                                        & (d,i)\in F \cup I
\end{array}
\end{array}$
\makeatcode
\caption{Total ILP formulation}\label{form:final}
\end{figure}

\section{Evaluation}\label{sec:benchmarks}
In this section, we examine the difference between our sample cost models, and see whether there is a large difference in performance between such a cost model or a greedy clustering algorithm.
We compare the performance by measuring the runtime of compiled Accelerate programs on a 16-core AMD Ryzen Threadripper 2950X.


\paragraph{Greedy}
We use two greedy algorithms in this study; one that prioritises fusing edges at the \emph{top} of the data dependence graph and one that prioritises fusing edges at the \emph{bottom}.
Both of these iteratively pick edges, try to fuse them, and check whether there is still a solution using our model.
The only difference is the order in which they traverse the list of edges.
Because they use an ILP solver to check for the existance of a solution at each iteration, they are not fast, but the solutions they find are always locally maximal (extra edges can only be fused if one first unfuses other edges).
Both algorithms are clearly not optimal, but as an overestimation of typical greedy fusion algorithms they are the best comparison we have.
Actual fusion algorithms commonly used in compilers usually do not support diagonal fusion, but these do.

\definecolor{c1}{RGB}{243,150,94}
\definecolor{c2}{RGB}{170,21,85}
\definecolor{c3}{RGB}{36,167,147}
\definecolor{c4}{RGB}{82,135,198}
\definecolor{c5}{RGB}{250,230,171}
\definecolor{c6}{RGB}{110,59,35}


\begin{figure}
  \begin{tikzpicture}
    \begin{axis}[
      bar width = 6pt,
      enlarge x limits = 0.16,
      width = .5\textwidth,
      ybar=2.7pt,
      ymin=0,
      ymax=5,
      legend style={at={(1,1)},
        anchor=north east,legend columns=1,draw=none,fill=none},
      legend cell align={left},
      legend image code/.code={%
        \draw[#1] (0cm,-0.1cm) rectangle (0.2cm,0.1cm);
      }, 
      ylabel={Normalised runtime},
      symbolic x coords={greedyTopDownBad, greedyBottomUpBad, FlashAttention, Multigrid, LULESH},
      xtick=data,
      xtick pos=left,
      visualization depends on={rawy \as \rawy},
      nodes near coords={\pgfmathprintnumber\rawy},
      restrict y to domain*={\pgfkeysvalueof{/pgfplots/ymin}:\pgfkeysvalueof{/pgfplots/ymax}},
      nodes near coords align={vertical},
      nodes near coords style={/pgf/number format/fixed, /pgf/number format/precision=1, /pgf/number format/fixed zerofill, font=\tiny, color=black},
      coordinate style/.condition={rawy > 10}{xshift=-10pt, yshift=-10pt},
      coordinate style/.condition={rawy > 25}{xshift=20pt},
      xticklabel style={
        font=\small,
        align=center,
        style={rotate=20}
        },
      ]

      \addplot[black,fill=c1,]
      coordinates {(greedyTopDownBad,1) (greedyBottomUpBad,1) (FlashAttention,1) (Multigrid,1) (LULESH,1) };
      \addplot[black,fill=c2]
      coordinates {(greedyTopDownBad,1.54294538198288) (greedyBottomUpBad,1.00708059290469) (FlashAttention,0.845647526879652) (Multigrid,0.905114261220461) (LULESH,0.827938107371296) };
      \addplot[black,fill=c3]
      coordinates {(greedyTopDownBad,1.00978296747171) (greedyBottomUpBad,19.6438368583689) (FlashAttention,0.831861636007883) (Multigrid,0.992425160848971) (LULESH,0.837543549025881) };
      \addplot[black,fill=c4]
      coordinates {(greedyTopDownBad,4.0889465854366) (greedyBottomUpBad,37.1991904624318) (FlashAttention,3.76622532842107) (Multigrid,2.47863220720633) (LULESH,2.61163331185919) };

      \legend{ILP,greedy $\downarrow$, greedy $\uparrow$, no fusion}
    \end{axis}
  \end{tikzpicture}
  \caption{Comparing ILP to greedy and no fusion}\label{fig:bench1}
\end{figure}
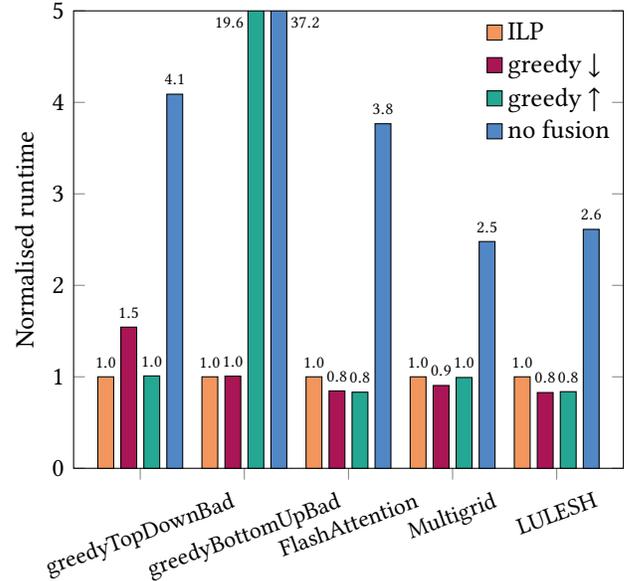

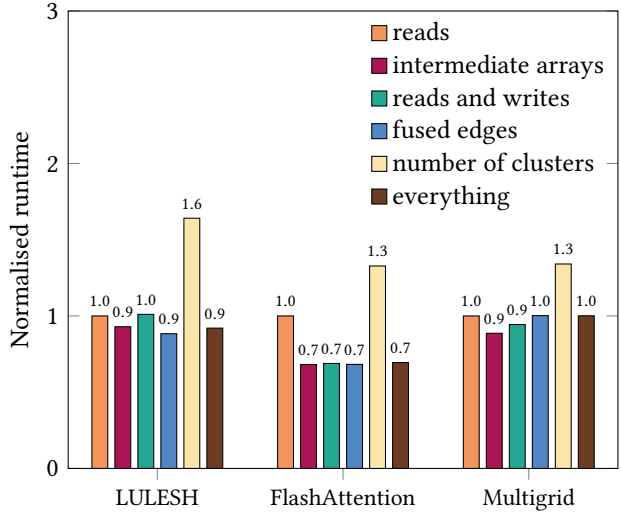
\begin{figure}
  \begin{tikzpicture}
    \begin{axis}[
      bar width = 6pt,
      enlarge x limits=0.24,
      width = .5\textwidth,
      ybar=2.7pt,
      ymax=3,
      ymin=0,
      legend style={at={(1,1)},
        anchor=north east,legend columns=1,draw=none,fill=none},
      legend cell align={left},
      legend image code/.code={%
        \draw[#1] (0cm,-0.1cm) rectangle (0.2cm,0.1cm);
      }, 
      ylabel={Normalised runtime},
      symbolic x coords={,LULESH, FlashAttention, Multigrid,},
      xtick=data,
      xtick pos=left,
      nodes near coords,
      nodes near coords align={vertical},
      nodes near coords style={/pgf/number format/fixed, /pgf/number format/precision=1, /pgf/number format/fixed zerofill, font=\tiny},
      xticklabel style={
        font=\small,
        align=center
        },
      ]
      \addplot[black,fill=c1]
      coordinates {(FlashAttention,1) (Multigrid,1) (LULESH,1) };

      \addplot[black,fill=c2]
      coordinates {(FlashAttention,0.681190683887872) (Multigrid,0.886329045306388) (LULESH,0.929235213784258) };
      
      \addplot[black,fill=c3]
      coordinates {(FlashAttention,0.688360387967554) (Multigrid,0.943835384174659) (LULESH,1.01036791654502) };
      
      \addplot[black,fill=c4]
      coordinates {(FlashAttention,0.682336072071049) (Multigrid,1.00212693319299) (LULESH,0.883475206313214) };
      
      \addplot[black,fill=c5]
      coordinates {(FlashAttention,1.32774446195189) (Multigrid,1.34074685414164) (LULESH,1.64095822074677) };
      
      \addplot[black,fill=c6]
      coordinates {(FlashAttention,0.69419113386357) (Multigrid,1.00104786942769) (LULESH,0.920122531875014) };
      
      
      
      
      

      \legend{reads,intermediate arrays,reads and writes,fused edges,number of clusters,everything}
    \end{axis}
  \end{tikzpicture}
  \caption{Comparing cost functions}\label{fig:bench2}
\end{figure}

\paragraph{Greedy counterexamples}
We have two small examples that showcase a simple trap that the greedy algorithms fall into.
In the first example, we @gather@ a large intermediate array that can be contracted if fused with both of the @fold@ operations.
The bottom-up greedy algorithm first fuses the final @map@ into @ys@, after which the two @fold@s can no longer be fused into the same cluster, forcing @large@ to be manifest.
This causes our specific benchmark (where @m@ is one million) to be twenty times slower than the optimal fusion assignment.
Note that none of the cost functions we show in this paper take the size of the array into account, but our formulation does support arbitrary weights on any fusible edge or contractable array.
\begin{python}
def greedyBottomUpBad(xs,m):
  n = shape xs
  is = generate ((n,m),lamn m: n)
  large = gather (is, xs)
  ys = fold (+, large)
  zs = fold (*, large)
  result = map (lamy: y + zs!0, ys)
  return result
\end{python}
For the top-down greedy algorithm, we cannot force it to manifest a @gather@ed array as easily.
Instead, we show a program where the greedy algorithm chooses a \emph{diagonal} fusion of @cs@ with @bs@, whereas the ILP is able to \emph{vertically} fuse @cs@ with @es@.
This only causes it to take about 1.5 times as long as the ILP solution.
\begin{python}
def greedyTopDownBad(as):
  n = shape as
  bs = map (lama:a*2, as)
  cs = map (lamb:b+1, bs)
  ds = generate (n, lami:i+bs!0)
  es = zipWith (+, cs, ds)
  result = fold (+, es)
  return result
\end{python}

\paragraph{Large case studies}
We measured the performance of our ILP formulation and the resulting fusion assignments on a few large data-parallel problems.
As these are too large to manually extract an ILP or fuse, we are restricted to the programs that are already implemented in our language: 
Livermore Unstructured Lagrangian Explicit Shock Hydrodynamics (LULESH) is a hydrodynamics simulation program~\cite{lulesh}.
FlashAttention is an exact attention algorithm that utilises tiling~\cite{flashattention}.
Multigrid is a three dimensional benchmark from the NAS Parallel Benchmarks~\cite{nasparallel}. It approximates the solution to a specific discrete Poisson problem.

\subsection{Runtime analysis}
We analyse the runtime of a few programs when compiled using one of our cost models, a greedy algorithm, or no fusion at all. The results are in Figures~\ref{fig:bench1} and~\ref{fig:bench2}.
The greedy algorithms perform well on the selected large benchmarks, but the small examples highlight their pitfalls.

When comparing the various cost functions, we see that the number of clusters is not a great target for runtime optimization.
The other cost functions perform very similarly on most benchmarks, with the exception of FlashAttention.
Our implementation of FlashAttention has multiple instances of the pattern where an $n$-dimensional array is @gather@ed to $n+1$-dimensions, and the result is then reduced back to $n$ dimensions with a @fold@.
We presume that the large difference between the cost functions is due to this pattern, where only considering the number of array reads does not manage to eliminate the $n+1$ dimensional arrays.

The fact that the greedy algorithms match or even slightly outperform the ILP solutions on the large benchmarks is a signal that none of the cost functions we consider in this paper effectively models expected run time on our machine.
Our ILP formulation of the search space, and the implementation in Accelerate that we used for these benchmarks, enable future research into cost functions that accurately optimize for runtime or other targets on specific hardware.
One example would be to create a linear combination of a set of cost functions, and use auto tuning to find the coefficients of this combined cost function.

\begin{table}
  \caption{ILP solve time (`reads and writes' metric)}
  \begin{tabular}{llll}
  \toprule
  Benchmark & Combinators & Gurobi \\
  \midrule
  FlashAttention & 66   & \SI{9.77}{s}\\
  LULESH         & 51   & \SI{0.42}{s}\\
  Multigrid      & 99   & \SI{0.29}{s}\\
  \bottomrule
  \end{tabular}
  \label{fig:solvetime}
\end{table}

\subsection{ILP solve time}
The inherent trade-off we explore with this work involves solving an NP-hard problem at compile time. 
We used the open-source ILP solver CBC version 2.10.7~\cite{cbc2-10-7}, as well as the commercial ILP solver Gurobi version 9.1.2~\cite{gurobi}. 
The results for Gurobi are shown in Figure~\ref{fig:solvetime}, but CBC was unable to solve any of the three large benchmarks within the time limit of one hour.

To give a rough indication of the size of our benchmark programs, we've included the number of array combinators in the table. 
As we can see by comparing FlashAttention with Multigrid, the sheer number of combinators is not an accurate predictor of the difficulty to find the optimal fusion assignment. 

Gurobi is capable of solving our problems much faster than CBC is, on the same hardware. 
This is unsurprising; the open-source solvers are known to currently be much slower than the top commercial solvers~\cite{comparingsolvers}.
We hypothesize that this is because Gurobi recognizes some of the structure in our ILP that CBC does not.
For example, for any clustering, there is a partial order on the $\pi$ variables, but the exact numerical values of them are irrelevant.
There is also hope for contexts where commercial solvers are not available: 
CBC found the optimal solution to these problems in minutes, and spent the rest of its time failing to prove that this solution is optimal.
It is possible to use solutions that solvers find even when they are not optimal, because our ILP defines a search space wherein all solutions represent a viable partitioning.
We have not attempted to write an approximation algorithm for any of our cost functions ourselves, but running a solver with a reasonable time limit could be viable.

\section{Related Work}\label{sec:relatedw}
Many optimal fusion algorithms consider parallel loops in their problem setting, but none are able to fuse @gather@ as well as our method~\cite{ILP,amosILP,complexityfusion,typedfusion,graphpartit,maximizingparallelism}.
Our Integer Linear Programming formulation is based on the formulation by Megiddo and Sarkar~\cite{ILP}.
The original formulation allows each fusible edge to have an arbitrary weight, and minimises the sum of the weights of unfused edges.
Our method would support this too, but in our current cost functions we set each of these weights to one.
We are not the first to build upon this work:
Darte and Huard add the notion of \emph{loop shifting} to the formulation, 
supporting a program transformation that enables them to fuse more loops~\cite{darte2002new}.
Robinson et al.~\cite{amosILP} steps away from the explict imperative loop nests,
instead fusing a set of array combinators, similar to our approach.
Their main contribution is the support of size-changing operations like @filter@,
by incorporating \emph{iteration sizes} in the ILP.\@
They produce an ILP that is quadratic in the size of the program, whereas we are able to stay linear (depending on the cost function).
Additionally, their work is inherently restricted to \emph{sequential} loops.
It would be possible to combine our approaches for a setting that has both parallel and sequential loops,
supporting @filter@s in sequential loops and @gather@s in parallelisable ones.
The concept of such a heterogeneous setting has also been explored~\cite{ILP,complexityfusion}, 
with either a constraint or a cost function to prohibit or disincentivize sequentializing parallel loops or combinators.
Typed fusion~\cite{typedfusion} can be seen as a generalisation of this, as it supports an arbitrary number of types, but it does not support arbitrary weights on the fusible edges.


Another successful approach to finding optimal fusion assignments and related loop transformations is the polyhedral model~\cite{grosser2012polly,polyhedralapplicable}.
The polyhedral models can also be expressed in terms of an ILP, but there are some key differences with our work.
The polyhedral approach takes iteration transformations like tiling into account, but it can only support affine iteration spaces. 
In particular, this excludes @gather@: under the polyhedral model, a @gather@ will never fuse with the producer of the source array.
In contrast, our work is based on extracting more static information from the
higher-level combinators, which lets us fuse some loops that the traditional models can't.

\section{Conclusions and Future Work}
In this paper we describe an extension to existing ILP-based fusion algorithms, which focusses on the common data-parallel array combinators.
This formulation relies entirely on the static properties of the array combinators themselves.
For example, we do not infer any structure in manual index computations.
The biggest contribution is the encoding of the rules surrounding @gather@-like combinators in linear constraints, which require us to consider the \emph{order} in which each array is produced and consumed.

Our fusion method does not consider partionings which lead to a duplication of work. In approaches without diagonal
fusion, this generally prevents fusion of arrays used more than once altogether. In our approach, in contrast, it will
fuse. However, duplicating work is in some cases still worthwhile, since it may enable vertically fusing a combinator into two separate
clusters, and thereby reduce memory consumption.  We plan to extend to this work with consideration 
for work duplication. 

Finally, but perhaps most crucially, we have not attempted to find an optimal cost function for any hardware yet.
It is a strength of this method that we are able to support any metric, and it is possible that one should use a
different cost function depending on the hardware and the objective (i.e.\ speed, energy efficiency, memory
consumption). Nevertheless, our fusion system itself is useless without a good cost function.

\makeatchar 
\bibliography{Bibl}
\end{document}